# Role of $Bi^{3+}$ ion substitution on the piezocatalytic degradation performance of lead-free $BaTi_{0.89}Sn_{0.11}O_3$ at low vibrational energy

Salma TOUILI[a,b], Sara GHAZI[b], M'barek AMJOUD[b], Daoud MEZZANE[a,b], Hana URŠIČ[c], Zdravko KUTNJAK[c], Bouchra ASBANI[a], Mustapha JOUIAD[a], and Mimoun EL MARSSI[a,*]

a: LPMC, University of Picardie Jules Verne, Amiens 80039, France

b: IMED-Lab, Cadi Ayyad University, Marrakesh, 40000, Morocco

c: Jožef Stefan Institute, Jamova Cesta 39, Ljubljana, 1000, Slovenia

* Corresponding author: mimoun.elmarssi@u-picardie.fr

**Abstract**

Harnessing low ultrasonic vibration energy to initiate piezocatalytic-based reactions is increasingly becoming a significant focus due to the current environmental energy crisis. In this study, we report on the potential of heterovalent ion doping in improving piezocatalytic degradation of Rhodamine B dye (RhB) under the application of low-power ultrasonic vibration. For this purpose, bismuth ions $Bi^{3+}$ are substituted into the lead-free ferroelectric $BaTi_{0.89}Sn_{0.11}O_3$ ($BTSn_{11}$-xBi, x = 0, 0.02, 0.04) submicrocubes synthesized using sol-gel method. The structural, morphological, optical and piezocatalytic properties of the prepared materials appear to be significantly influenced by the substituting content. Compared to pure $BTSn_{11}$ and $BTSn_{11}$-0.04Bi, $BTSn_{11}$-0.02Bi exhibits the lowest band gap energy value of 3.22 eV, the smallest particle size value of 283 nm, the strongest piezoelectric current of about 8 $\mu A/cm^2$, and the lowest required applied field for piezoresponse force microscopy hysteresis loops. The degradation efficiency and the kinetic rate constant are substantially higher for $BTSn_{11}$-0.02Bi, indicating its enhanced piezocatalytic efficiency. Total organic carbon (TOC) measurements revealed good RhB mineralization using $BTSn_{11}$-xBi piezocatalysts. Furthermore, the $BTSn_{11}$-0.02Bi sample demonstrates good reusability and stability by maintaining effective RhB dye degradation for three consecutive cycling tests. These findings suggest that the $BTSn_{11}$-0.02Bi piezocatalytic system holds significant potential for environmental remediation applications.

## 1. Introduction

Combating and reducing water pollution has become an essential and urgent priority for protecting the environment and promoting sustainability. The dyeing and textile industries are responsible for 20% of the total volume of industrial wastewater [1,2]. In addition to environmental protection, addressing dye-containing wastewater is crucial for safeguarding human well-being [2]. In this respect, catalysts capable of transforming sustainable energy into chemical energy represent a promising solution to environmental pollution worldwide [3]. For this purpose, scientists have explored different energy conversion strategies adapted to each existing renewable energy source. Mechanical energy can be regarded as an abundant form of energy that can be harvested to drive the innovative piezocatalysis process [4]. Specifically, piezocatalysis can efficiently harness natural low-frequency mechanical energy such as water wave energy, wind, and tide, among others [4,5]. Furthermore, among its many other catalytic applications, piezocatalysis has demonstrated its effectiveness in treating a variety of water pollutants [6].

In the piezocatalysis process, the applied stress, such as ultrasonic vibration, can cause charge separation within non-centrosymmetrically structured piezoelectric materials, leading to an electric dipole polarization [7]. This induced polarization leads to the generation of a piezoelectric potential, which can promote intrinsic charge carrier separation, thereby enhancing catalytic properties [6]. Particularly, even low vibrational energy applied to piezoelectric nanomaterials, like that produced by muscle contraction, can produce the piezoelectric potential [8,9]. Through redox reactions with surrounding oxygen and water molecules, piezoelectric materials can produce reactive oxygen species (ROS) during the piezocatalysis process, such as superoxide anions and hydroxyl radicals which are the driving force behind the decomposition of organic pollutants. In piezocatalysis, ROS generation is closely related to ultrasonic vibration parameters, notably ultrasonic power and frequency. It is worth noting that a high ultrasonic frequency and power are required to trigger piezocatalytic reactions [4]. In this respect, Lan et al. [10] suggested that higher ultrasonic power or frequency (100 W, 40 kHz) can boost the piezo-photocatalytic performance of La-doped $BiFeO_3$ by increasing catalyst stress, thereby enhancing the piezoelectric field generated. However, while increasing piezocatalytic process efficiency is valuable, developing systems with less energy consumption is crucial to harness low-vibration energy sources and tackling the energy crisis in water treatment. Moreover, high ultrasonic energy can induce excessive heating effects, which, in turn, may lead to changes in the physical properties of the piezocatalytic system concerned. In particular, the temperature rise can induce the evaporation of dissolved oxygen needed to generate the superoxide ions required to degrade water contaminants.

Currently, ferroelectric $BaTiO_3$-based solid solutions typically receive greater attention due to their perovskite structure's diverse properties [11]. Wu et al. [12] found that $BaTiO_3$ nanowires could effectively degrade the pollutant dye methyl orange (MO) under ultrasonic vibration with a power of 80 W. Specifically, the lead-free ceramic $BaTi_{0.89}Sn_{0.11}O_3$ ($BTSn_{11}$) showed a Curie temperature ($T_c$) close to room temperature and an ultra-high piezoelectric coefficient $d_{33}$ > 1100 pC/N [13]. Additionally, it has demonstrated enhanced piezoelectric response when embedded in a PVDF-HFP polymer matrix and improved dielectric response near room temperature [14,15]. Furthermore, Zhao et al. [16] presented compelling proof of an optimum working temperature close to $T_c$ for $BTSn_{11}$ nanoparticles subjected to ultrasonic vibration and thermal variation. However, practical applications of piezocatalysis are generally limited due to the low piezocatalytic reaction rates [3]. For this reason, the ion doping strategy can be considered an important method for tailoring piezocatalytic properties. In principle, the incorporation of elements in the A or B sites can theoretically lead to atomic interactions that cause lattice distortion and symmetry reduction, which in turn can improve the piezoelectric properties of the ABO3-type structure [17]. For instance, You et al. [18] demonstrated remarkable piezocatalytic degradation performance by Ba-substituted $SrTiO_3$ crystals using a high ultrasonic power of 100 W. Alternatively, Yu et al. [3] suggested an effective heterovalent ion doping strategy by selecting the $Li^+$ acceptor ion and $La^{3+}$ donor ion doped separately in the $BaTiO_3$ matrix to enhance piezocatalytic activity towards specific pollutants under different ultrasonic vibration power (60, 80, 100 W) and frequency (20, 40, 60 kHz). As expected, improved piezocatalytic activity was achieved at appropriate doping amounts and high ultrasonic power conditions. In fact, donor doping is a powerful strategy for substantially enhancing the piezoelectric properties and remanent polarization of ferroelectric materials [19]. Meanwhile, the effect of heterovalent $Bi^{3+}$ substitution on the ferroelectric and piezoelectric properties of $BaTiO_3$ and its derivative solid solutions has been extensively studied [20,21]. Particularly, the active $6sp^2$ lone pair of $Bi^{3+}$ donor cations can induce local structural distortion in $BaTiO_3$-based materials, thereby increasing their ferroelectric polarization [22]. The $Bi^{3+}$ substitution can also improve the optical and photocatalytic activity of perovskite based materials due to its direct influence in their band alignment [23].

In this study, we have explored the role of $Bi^{3+}$ heterovalent substitution on the piezocatalytic activity of lead-free $BTSn_{11}$ for RhB dye degradation. In order to reduce the piezocatalytic energy consumption, the piezo-degradation of RhB was carried out under low-power ultrasonic vibration of 30 W. The $BTSn_{11}$-xBi powders were synthesized using a sol-gel method, and their structure, morphology, optical characteristics, piezoelectric properties, RhB-degradation performance, Total Organic Carbon (TOC) measurements, and reactive species trapping experience have been thoroughly characterized.

## 2. Experimental details

### 2.1. Synthesis of $BTSn_{11}$-xBi

Pure $BaTi_{0.89}Sn_{0.11}O_3$ ($BTSn_{11}$) and Bi substituted $BaTi_{0.89}Sn_{0.11}O_3$ ($BTSn_{11}$-xBi) powders with the general formula $Ba_{1-x}Bi_{2x/3}Ti_{0.89}Sn_{0.11}O_3$ (with x = 0.00, 0.02, 0.04) were prepared by a sol-gel method. Appropriate amounts of barium acetate ($Ba(CH_3COO)_2$) and bismuth nitrate ($Bi(NO_3)_3 . 5H_2O$) were dissolved in glacial acetic acid ($CH_3COOH$) with the proper stoichiometric ratio (solution 1), while tin chloride and titanium isopropoxide ($C_{12}H_{28}O_4$-Ti) were dissolved in 2-methoxyethanol ($C_3H_8O_2$) (solution 2) and stirred for 1h. Then, solutions 1 and 2 were mixed at room temperature. After stirring the mixture solution for 1h, ammonia ($NH_4OH$) was carefully dropped until it became a transparent and clear gel. The gel was dried overnight at 80°C, then calcined at 1000°C for 4h to get fine $BTSn_{11}$-xBi powders.

To analyze the optical properties of the samples, $BTSn_{11}$-xBi films were prepared using the spray-coating technique [24]. To this end, each sample solution with a molar concentration of 0.1 M in ethanol solvent was ultrasonically dispersed and deposited at a 1 ml/min rate onto the ITO substrate at 350°C.

### 2.2. Characterisation

The structure of $BTSn_{11}$-xBi powders was characterized at room temperature by adopting Cu Kα radiation (λ = 1.54059 Å) on a Panalytical X-Pert Pro X-rays diffraction analyser ranging from 20° to 80°. Raman spectra of powder samples were carried out at room temperature in the wavenumber range of 100 $cm^{-1}$–800 $cm^{-1}$ with green excitation laser of 532 nm using a Renishaw in Via Reflex Raman spectrometer equipped with an Edge-filter. The microstructure of the prepared powders was analyzed by scanning electron microscope (SEM) (Environmental Quanta 200 FEG, FEI). A UV-visible-near IR spectrophotometer (JASCO-60) running in transmission mode was employed to investigate the optical properties of $BTSn_{11}$-xBi films prepared by the spray coating technique in the 340-800 nm spectral range.

The transient piezocurrent response was measured under the application of a low ultrasonic vibration of (35 kHz, 30 W) using a conventional three-electrode cell assembly connected to an electrochemical workstation (VSP-3e). The three-electrode cell consists of Ag/AgCl as the reference electrode, Pt wire as the counter electrode, and $BTSn_{11}$-xBi films deposited on an ITO substrate as the working electrode. Sodium sulfate (0.5 M, pH = 7) was used as the electrolyte solution.

AFM topography and piezoresponse force microscopy (PFM) analyses of prepared samples were performed using an MFP-3D atomic force microscope (AFM, Asylum Research, Santa Barbara, California, USA). To avoid the powder sticking to the PFM tip and to ensure a flat surface for AFM scanning, the powders were fixed using epoxy resin, as previously reported in ref. [25]. The samples were then polished through metallographic procedures explained in detail in ref. [26]. Ti/Pt-coated Si tips on Al-coated Si cantilevers (OMCL-AC240TM-R3, Olympus, Japan) with a radius of 25 nm were used. PFM out-of-plane amplitude images were acquired with a PFM dual AC resonance tracking (DART) switching spectroscopy (SS) mode by applying an AC voltage with an amplitude of 10 V and a frequency of 340 kHz. The local hysteresis loops were also measured with PFM, as described in ref. [26]. The DC electric fields consisted of a series of increasing steps with a frequency of 20 Hz and maximum amplitudes ranging from 50 V to 150 V. This was complemented by a triangular envelope with a frequency of 0.2 Hz. Additionally, a superimposed sinusoidal AC signal with an amplitude of either 5 or 10 V and a frequency of about 350 kHz was used. Three cycles were measured in an off-electric field mode.

UV-vis diffuse spectra for piezo-degradation analysis were measured using a Shimadzu UV-2600 spectrophotometer for wavelengths ranging from 450 to 650 nm.

### 2.3. Piezocatalytic activity

The piezocatalytic degradation performance of the prepared $BTSn_{11}$-xBi powders (x = 0, 0.02, 0.04) was investigated using 20 ml of 5 mg/L concentrated Rhodamine B dye and 20 mg of piezocatalyst under low ultrasonic energy (35 kHz, 30 W) in dark conditions. Before starting the piezocatalytic degradation experiments, the samples and the RhB dye solution were stirred in the dark for 60 minutes to achieve the adsorption-desorption equilibrium. Hydrogen peroxide (30 % by weight) was added to the mixture to kick-start the catalytic reaction. Afterward, the mixture solution was transferred to the ultrasonic cleaner. 2 ml of RhB dye was sampled at regular intervals and separated from the catalyst by centrifugation. In addition, to prevent heat degradation, the ultrasonic bath water was changed every 30 min.

The degradation efficiency (η) was calculated from the concentration of the highest absorption peak ($\lambda$ = 554 nm) based on the following equation:

$$\eta(\%) = (C_0-C_t)/C_0 \times 100 \qquad (1)$$

where, '$C_t$' and '$C_0$' are the concentrations of RhB dye at time = 't' (min) and at a time 't' = 0 (min). A UV–Vis near IR spectroscope was used to determine the solution's concentration at RhB maximal absorption peaks.

To assess the degree of mineralization of the piezocatalytic degradation of the RhB organic dye, the Total Organic Carbon (TOC) of residual organic dye and any by-products was measured after the test using a TOC analyzer (TOC-L CPH/CPN, Shimadzu). Furthermore, to investigate the piezocatalytic mechanism underlying RhB degradation, isopropanol (IPA) was used to trap hydroxyl radicals ($^{\bullet}OH$) [27], methanol was used as a hole ($h^+$) scavenger [28], and $N_2$ was served as a quencher of superoxide radical anions ($^{\bullet}O_2^-$) [29]. Before running the piezocatalytic experiment, the appropriate amount of scavenger was added to the dye-catalyst mixture.

## 3. Results and discussions

### 3.1. Structural and microstructural analysis

The XRD patterns of $BTSn_{11}$ with various $Bi^{3+}$ ions concentrations are depicted in Fig. 1(a). The observed patterns demonstrate the existence of a single perovskite structure without the appearance of a secondary phase at the sensitivity limit of the device's detection, revealing the effective diffusion of $Bi^{3+}$ ions into the $BTSn_{11}$ crystal lattice. Fig. 1(b), (c) illustrates the enlarged view of XRD patterns in the 2θ range of 44–46° and 64–67°. It is widely recognized that tetragonal and cubic phases can be distinguished by analyzing the (002)/(200) peaks that appear in the 44–46° 2θ range [15]. For pristine $BTSn_{11}$, the peak (002) shows a distinct splitting, indicating the existence of a tetragonal phase with a P4mm space group [15]. When the Bi doping concentration changes, the tetragonal peaks (002) and (200) tend to merge into a single peak, suggesting the formation of a pseudocubic phase [30]. Otherwise, the broad appearance of the (002)/(200) peaks, particularly observed for $BTSn_{11}$-0.04Bi, could be associated with their smaller crystallite size and the presence of microstrain [31]. Similar results were previously demonstrated by Kumar et al. by doping $La^{3+}$ heterovalent ions on $Ba(Ti_{0.95}Sn_{0.05})O_3$ material [32]. Furthermore, with increasing Bi content in the $BTSn_{11}$ lattice, a shift of the diffraction plane (200) towards the higher angles is observed, suggesting a decrease in the lattice parameter and, thus, in the volume of the unit cell, which indicates the deformation of $BTSn_{11}$ lattice by the penetration of the $Bi^{3+}$ ions and also the generation of oxygen vacancies [21]. A similar result was also reported by Zhou et al. [33]. The lattice shrinkage is ascribed to the substitution of $Bi^{3+}$ ions whose ionic radii (1.36 Å) are relatively smaller than those of $Ba^{2+}$ (1.61 Å). Furthermore, to maintain charge balance, the $Bi^{3+}$ ions in the $BTSn_{11}$ lattice's A sites carry positive charges that can be compensated by electrons and/or barium vacancies [33]. Hence, the substitution of $Bi^{3+}$ for $Ba^{2+}$ will take place in the following way [33]:

$$Bi_{Ba} \Leftrightarrow Bi_{Ba}^{\cdot} + e' \quad \text{(i)}$$

$$Bi_{Ba} \Leftrightarrow Bi_{Ba}^{\cdot} + \frac{1}{2} V_{Ba}'' \quad \text{(ii)}$$

where $Bi_{Ba}^{\cdot}$ represents an ionized Bi donor, $V_{Ba}''$ is a doubly ionized barium vacancy, and e' an electron [33]. Therefore, bismuth substitution directly causes a barium vacancy through a charge compensation mechanism [34]. The creation of A-site vacancies in the $BTSn_{11}$ matrix leads to compositional fluctuations and lattice distortions [35].

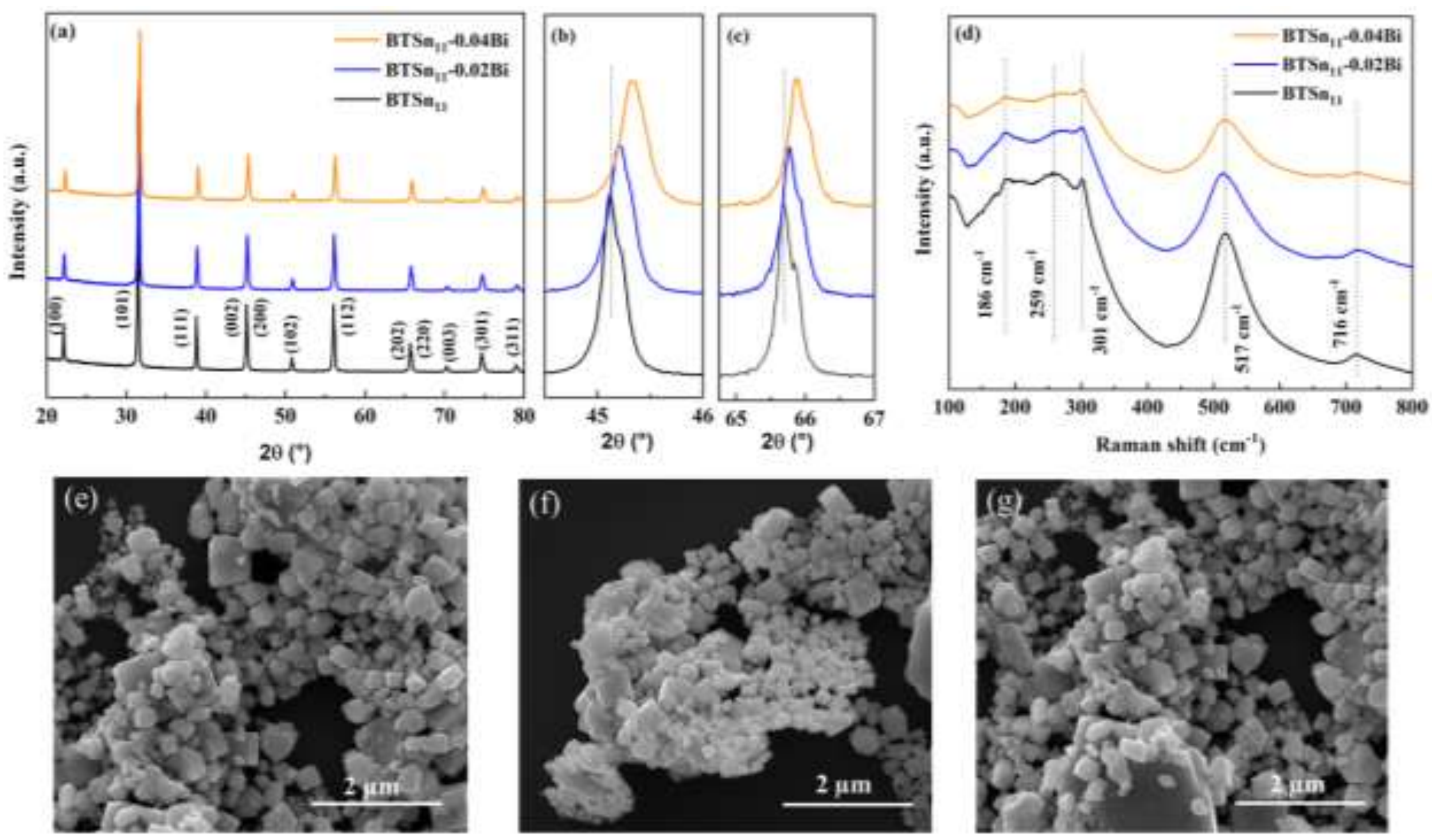

Fig. 1. (a) X-ray patterns of $BTSn_{11}$-xBi (x = 0, 0.02, 0.04) powders, (b) and (c) enlarged views of XRD peaks centered around 45.5° and 65.5°, respectively, (d), Raman spectra of $BTSn_{11}$-xBi powders, (e), (f) and (g) SEM images of $BTSn_{11}$, $BTSn_{11}$-0.02Bi and $BTSn_{11}$-0.04Bi powders, respectively.

Table 1 lists the structural properties of the $BTSn_{11}$-xBi samples obtained according to the XRD spectra. These properties include lattice parameters and the unit cell volume calculated by Unitcell software, as well as crystallite size and lattice strain determined by the Debye Scherrer formula. It should be noted that with increasing $Bi^{3+}$ ions concentrations, a decrease in $BTSn_{11}$-xBi (x = 0, 0.02, 0.04) crystallite size is observed. The crystallite size value reached its minimum (38.9 nm) when doped with 4% Bi, corresponding to the greatest lattice strain. A decrease in crystallite size matched the upward trend in lattice strain. Notably, the X-ray diffraction peaks broadening is primarily due to lattice strain caused by crystallite defects and distortions [36]. Consequently, the significant lattice strain of the $BTSn_{11}$-0.04Bi sample is behind the considerable broadening of its diffraction peaks.

Table 1. Lattice parameters, cell volume, crystallite size, lattice strain, and particle size of $BTSn_{11}$-xBi powders.

| x | Lattice parameters | | | Cell volume | Crystallite | Lattice strain | Particle size (nm) |
|---|---|---|---|---|---|---|---|
| | a (Å) | b (Å) | c (Å) | V (Å$^3$) | size (nm) | ($\times10^{-3}$) | |
| 0.00 | 4.0103 | 4.0103 | 4.0146 | 64.5663 | 53.1 | 1.3 | 330 |
| 0.02 | 4.0109 | 4.0109 | 4.0110 | 64.5265 | 42.2 | 1.7 | 283 |
| 0.04 | 4.0003 | 4.0003 | 4.0036 | 64.0663 | 38.9 | 2.1 | 320 |

Since the structural information from XRD analysis is based on a group of unit cells larger than 10 $nm^3$, its experimental accuracy in identifying the mixed phases of materials is limited. Accordingly, Raman spectroscopy, a versatile method to probe the short-range modes since its structural information refers to a single $TiO_6$ octahedron, was used [32]. Fig. 1(d) shows the Raman spectra of $BTSn_{11}$ and $BTSn_{11}$-xBi samples at room temperature in the 100–800 $cm^{-1}$ wavenumber range. The Raman spectra show five modes at 186, 259, 301, 517, and 716 $cm^{-1}$. The dip observed at 186 $cm^{-1}$ is attributed to the $A_1(TO_2)$ anti-symmetry mode [11]. The bands centered at 259 and 517 $cm^{-1}$ are associated with the transverse optical modes of $A_1$ symmetry [37], while the band at 301 $cm^{-1}$ is assigned to the $B_1$ mode. Lastly, the broadband at 716 $cm^{-1}$ is a consequence of the $A_1$ $(LO_3)$E (LO) mode [38]. The tetragonal phase can be identified by examining the relative intensity of the bands positioned at 301 and 716 $cm^{-1}$ [7]. Introducing $Bi^{3+}$ ions into the $BTSn_{11}$ matrix affects the Raman bands intensities. As the amount of Bi substitute increases, the band intensity of $BTSn_{11}$-xBi gradually decreases throughout the range of Raman spectra. The 301 and 716 $cm^{-1}$ bands are prominent for pristine $BTSn_{11}$, although they broaden with progressively decreasing intensity for $BTSn_{11}$-0.02Bi and further for $BTSn_{11}$-0.04Bi, which may indicate the first signs of a phase change from tetragonal to cubic [38]. Furthermore, the 517 $cm^{-1}$ bands broadening with increasing doping indicates a local crystal structure disorder. This reflects the modification of the crystalline symmetries induced by the compositional change. Similarly, Ju et al. [30] found that the tetragonal structure gradually disappeared with increasing $Cr^{3+}$ doping concentration in the $BaTiO_3$ matrix, as indicated by reduced Raman peak intensities and increased peak broadenings. It is worth noting that the phase transition in doped nanocrystals is a complex interaction between order-disorder and displacement mechanisms [30]. Hence, based on the XRD and Raman findings, it can be deduced that $Bi^{3+}$ cations are successfully substituted in the $BTSn_{11}$ matrix and cause more structural disordering.

Scanning electron microscopy (SEM) was performed to evaluate the morphology of the $BTSn_{11}$-xBi particles. For this purpose, the sample powders were dispersed in ethanol, after which a few drops of the resulting solution were subsequently applied to the silicon substrate. The SEM micrographs are shown in Fig. 1(e)–(f). The $BTSn_{11}$-xBi powders mainly featured submicrometric cubes, and their size depended on the Bi doping content. As presented in Table 1, the average particle size of $BTSn_{11}$, $BTSn_{11}$-0.02Bi, and $BTSn_{11}$-0.04Bi is 330 nm, 283 nm, and 320 nm, respectively. The reduction of particle size of $BTSn_{11}$-0.02Bi suggests that a low concentration of Bi doping can effectively control particle size. The slight increase in particle size observed in $BTSn_{11}$-0.04Bi could be due to the structural change occurring in the $BTSn_{11}$ matrix by increasing Bi content, as described above.

### 3.2. Optical and piezo-electrochemical properties

The absorbance spectra of the prepared $BTSn_{11}$-xBi samples in the range of 340–800 nm are depicted in Fig. 2(a). The spray coating method was used to deposit $BTSn_{11}$-xBi submicrocubes on a glass substrate to assess their optical performances. All samples appeared to have a sharp absorbance of light below 380 nm, whereas less energy was absorbed for the rest of the spectrum. The energy bandgap ($E_g$) values of the $BTSn_{11}$-xBi samples were determined through Tauc's relation by the following formula [11]:

$$\alpha(h\nu) = B\,(h\nu - Eg)^{m} \qquad (2)$$

Where α refers to the coefficient of the absorption, B represents the energy independent coefficient, Eg is the energy bandgap of the prepared samples, h depicts the plank's constant, ν characterizes the frequency of the light, and m is the nature of the electronic transition causing optical absorption. As previously reported, $BTSn_{11}$ is a direct semiconductor; m is taken as 2 [3]. Figure 2(b) shows the predicted $BTSn_{11}$-xBi energy bandgaps based on Taucs' plot. The band gap energies of $BTSn_{11}$, $BTSn_{11}$-0.02Bi, and $BTSn_{11}$-0.04Bi are 3.31 eV, 3.22 eV, and 3.29 eV, respectively (Fig. 2(b)). The calculated Eg of sol-gel synthesized $BTSn_{11}$ submicrocubes showed the same value as the Eg of hydrothermally synthesized $BTSn_{11}$ nanoparticles (3.31 eV) [19]. For the bismuth-doped $BTSn_{11}$ samples, the bandgap decreases to 3.22 eV for $BTSn_{11}$-0.02Bi, then gradually increases to 3.29 eV for $BTSn_{11}$-0.04Bi. On the one hand, the shift in $E_g$ towards lower energy values can be explained by the structural disorder in the lattice due to the creation of A-site vacancies, oxygen vacancies, and deformations in the [$TiO_6$] clusters [40]. On the other hand, the observed shift towards the higher energy side when adding 4% of $B^{3+}$ ions can be explained by the Burstein-Moss effect [41,42]. According to this effect, the bandgap broadening mainly originated from the occupation of the electrons arising from the oxygen vacancies created by the excessive doping defects in the lowest states near the edge of the conduction band [43]. A similar trend of Eg change was observed in Zr-doped $BaTiO_3$ thin film [43].

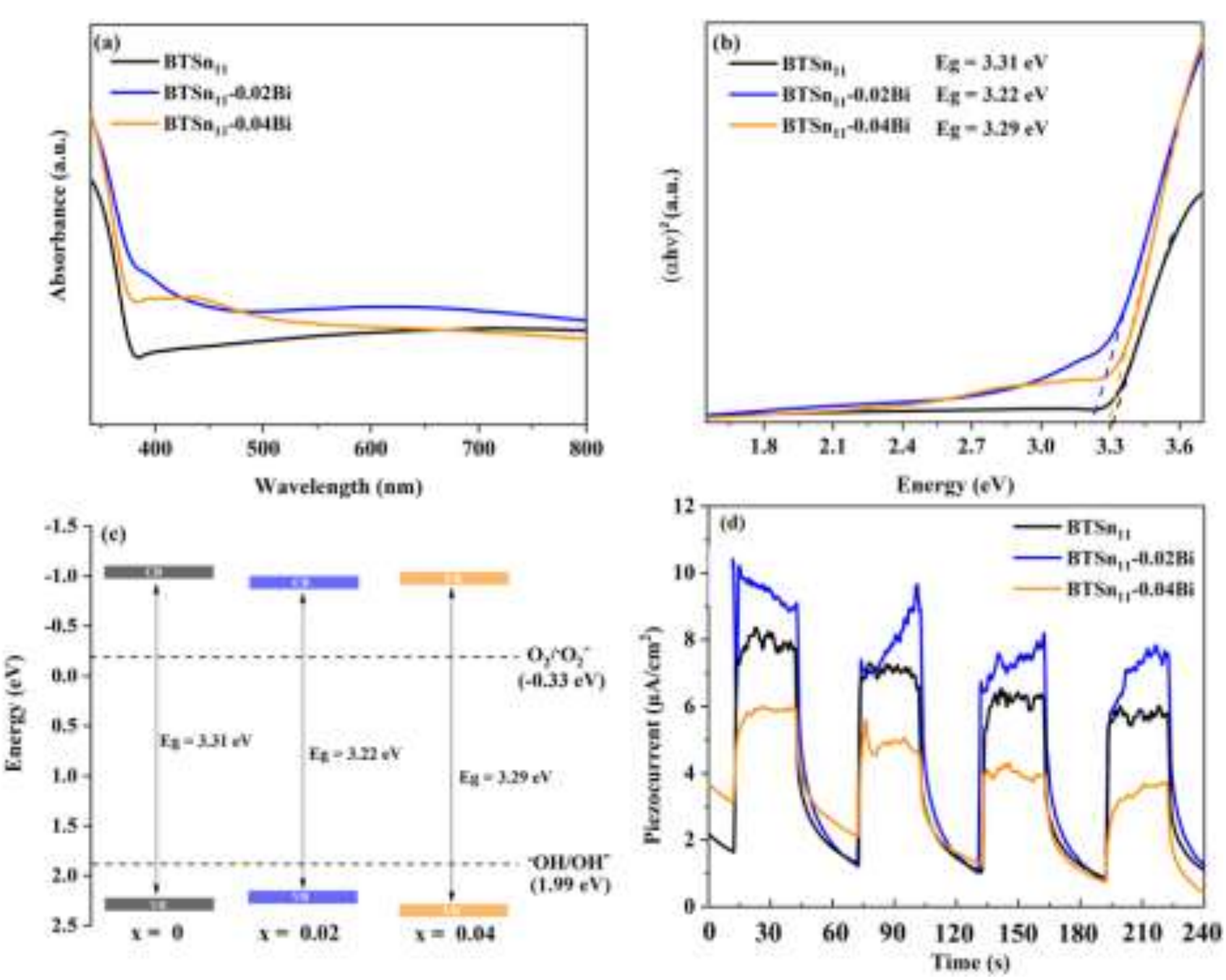

Fig. 2. (a) UV-vis DRS spectra of $BTSn_{11}$-xBi samples, (b) corresponding plot $(\alpha h\nu)^2$ versus energy $h\nu$, (c) schematic representations of the calculated CB and VB for $BTSn_{11}$-xBi samples, (d) transient piezo-current density recorded at 1 V vs Ag/AgCl under on-off ultrasonic vibration.

Moreover, the conduction band position and the valence band position of the $BTSn_{11}$-xBi samples can be calculated through the following Mulliken electronegativity empirical equations:

$$E_{VB} = \chi - E_0 + 0.5E_g \quad (3)$$

$$E_{CB} = E_{VB} - E_g \quad (4)$$

Where $E_{VB}$ and $E_{CB}$ are the valence band and conduction band potentials, respectively, $E_0$ represents the energy of the free electrons versus hydrogen, which equals to 4.5, and $\chi$ is the electronegativity of the semiconductor, which can be calculated following the given equation [36]:

$$\chi = [\chi(A)^a \chi(B)^b \chi(C)^c \chi(D)^d \chi(E)^e]^{1/a+b+c+d+e} \quad (5)$$

The parameters a, b, c, d, and e refer to the number of atoms in each composition, and $\chi$ is the electronegativity of each element in the component in eV, which was derived in this case from Pearson's research work [44]. The obtained $\chi$ values for $BTSn_{11}$, $BTSn_{11}$-0.02Bi, and $BTSn_{11}$-0.04Bi samples are 5.154 eV, 5.168 eV, and 5.181 eV, respectively. The calculated $E_{CB}$ and $E_{VB}$ edge potentials for $BTSn_{11}$, $BTSn_{11}$-0.02Bi, and $BTSn_{11}$-0.04Bi are as follows: (-1.001 eV, 2.309 eV), (-0.942 eV, 2.278 eV), and (-0.964 eV, 2.326 eV), respectively. The schematic representation of these potentials for the $BTSn_{11}$-xBi samples is depicted in Fig. 2(c). It can be observed that when the degree of substitution increases, both $E_{CB}$ and $E_{VB}$ are affected. Therefore, $Bi^{3+}$ doping allows for controlling the energy required to excite electrons from the valence band to the conduction band.

The charge transfer capacity of the piezocatalyst primarily controls the kinetics of the dye degradation process [45]. For this purpose, the piezoelectric current measurements were conducted for each sample under 30 W ultrasonic vibration. Chronoamperometric measurements were recorded cyclically using the prepared working electrodes during a 30-second interval. Fig. 2(d) represents the vibration time-dependent piezoelectric current response for $BTSn_{11}$-xBi samples. The maximum current density is obtained for $BTSn_{11}$-0.02Bi, reaching 8 μA/cm$^2$, which is greater than previously reported values for typical piezocatalysits such as ZnO (0.15 μA/cm$^2$) [46] and $Bi_2WO_6$ nanosheets (0.3 μA/cm$^2$) [47]. This result indicates efficient charge transfer and separation of piezoelectrically generated carriers. This strong piezoelectric current signal is related to forming a built-in electric field by the applied ultrasonic vibration, which facilitates charge carrier separation [48]. The performance evolutions are directly related to the structural and microstructural changes. Therefore, the $BTSn_{11}$-0.04Bi piezoelectric current behavior originates from the trapping of electrons by excessive defects, resulting in a relatively low current density compared to pristine $BTSn_{11}$ and $BTSn_{11}$-0.02Bi. Moreover, the detected piezo-current is a result of the vibration-induced electric potential, and this potential is ultimately responsible for generating charge carriers [49,50].

### 3.3. Piezoelectric response of $BTSn_{11}$-xBi powders

Piezo-response force microscopy (PFM) measurements are crucial for characterizing the piezoelectric behavior of ferroelectric powders in the nanometer range. The local piezoelectric properties of $BTSn_{11}$-xBi ferroelectric catalysts were evaluated by PFM analysis. The AFM topography height image and piezoelectric amplitude and phase images of $BTSn_{11}$, $BTSn_{11}$-0.02Bi and $BTSn_{11}$-0.04Bi powders embedded in the epoxy matrix are shown in Fig. 3. In the PFM amplitude images (panels (b), (f) and (j)), the bright contrast of the $BTSn_{11}$-xBi powders shows their piezoelectric activity in contrast to the dark contrast of the non-piezoelectric epoxy. A PFM switching spectroscopy experiment was also used to determine the local PFM amplitude and phase hysteresis loops. These hysteresis loops show a typical ferroelectric/piezoelectric behavior and thus confirm the ferroelectric and piezoelectric behavior of all three prepared powders. The PFM hysteresis loops of the $BTSn_{11}$-0.02Bi powders were obtained at the lowest applied field, namely at a maximum amplitude of 50 V for the DC signal and an amplitude of only 5 V for the superimposed sinusoidal AC signal. In contrast, a higher electric field was required to obtain the hysteresis loops in the $BTSn_{11}$ and $BTSn_{11}$-0.04Bi powders. This was done with a maximum DC amplitude of 150 V and 80 V and amplitude of the superimposed sinusoidal AC signal of 10 V.

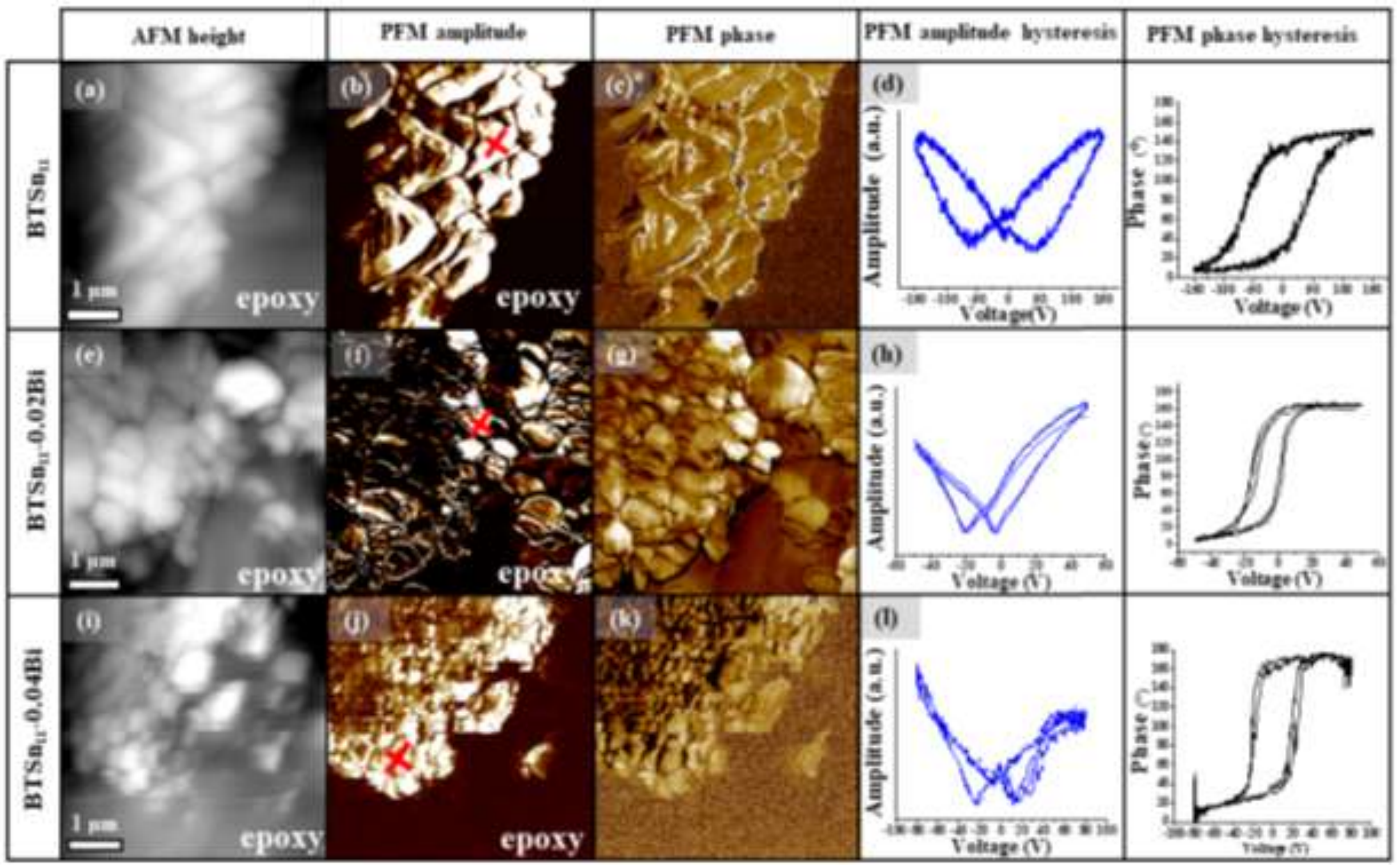

Fig. 3. (a), (e), (i) AFM topography heigh images. PFM (b), (f), (j) amplitude and (c), (g), (k) phase images. (d), (h), (l) PFM amplitude and phase hysteresis loops of $BTSn_{11}$-xBi powders embedded in the epoxy matrix. Local hysteresis loops were measured in the spot marked by a red cross in panels (b), (f), (j).

### 3.4. Piezocatalytic RhB degradation

Piezocatalytic activity of $BTSn_{11}$-xBi samples was evaluated by degrading RhB solution using low-energy ultrasonic vibration (35 kHz, 30 W). Fig. 4 illustrates the time evolution of the piezo-degradation of RhB dye in the presence of $BTSn_{11}$-xBi piezocatalysts. RhB piezocatalytic degradation was carried out for 300 min, and the absorption spectra were recorded after every 30 min of vibration. The intensity of the 554 nm maximum absorption peak of RhB reduces with increasing vibration time, revealing its gradual degradation. The RhB degradation trend of the adopted samples is the same as that of their bandgap evolution, whereas BTSn-0.02Bi, with the lowest bandgap value, degrades RhB more efficiently. Particularly, the intensity of the absorption peak rapidly decreases in the presence of $BTSn_{11}$-0.02Bi piezocatalyst, proving its excellent piezocatalytic degradation performance by harvesting vibration energy.

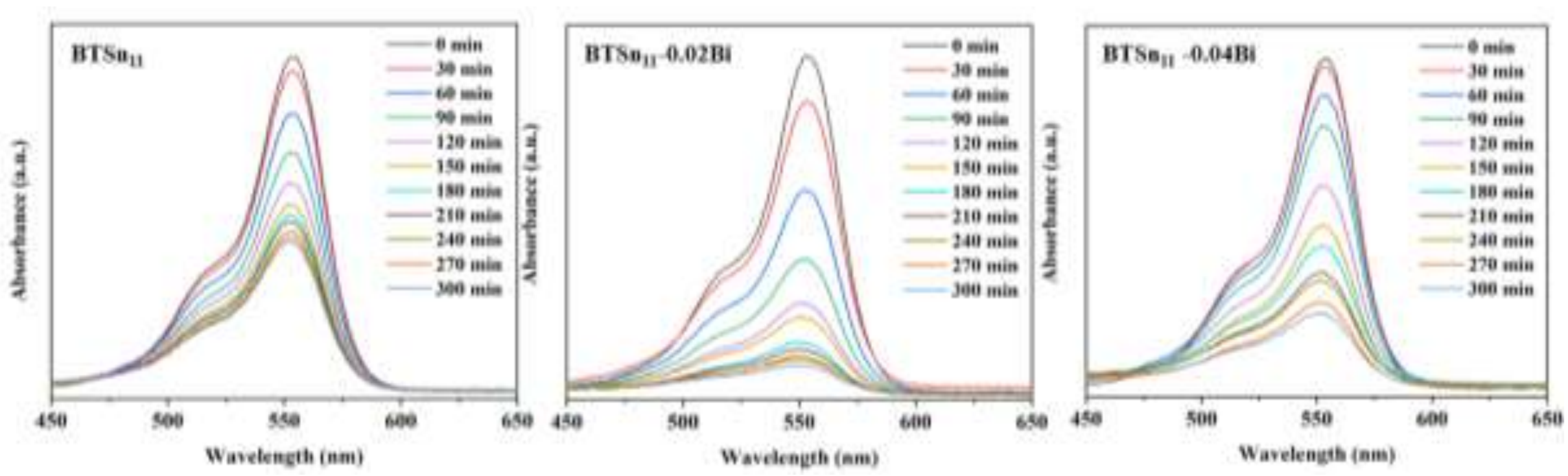

Fig. 4. UV–Vis's absorption spectra of RhB dye solution separated from $BTSn_{11}$-xBi catalyst suspensions.

For clarity of description, the RhB absorption peaks presented by the studied samples were transformed into $C_t/C_0$ lines, as depicted in Fig. 5(a). The blank test demonstrates that after 300 min of ultrasonic vibration, the degradation of the RhB dye can be neglected, suggesting that ultrasonic vibration cannot decompose RhB without the presence of the piezocatalysts. Furthermore, the degradation of RhB by undoped $BTSn_{11}$ is only 55%, whereas with the introduction of 2% $Bi^{3+}$ ions, it improves further to reach 92% and 79% for $BTSn_{11}$-0.04Bi for 300 min. In order to conduct a more methodical comparison of the piezocatalytic reaction rate (k) of the samples, a pseudo-first-order kinetic model is applied, as shown in the following equation :

$$-\ln(C/C_0) = k \, . \, t \qquad (6)$$

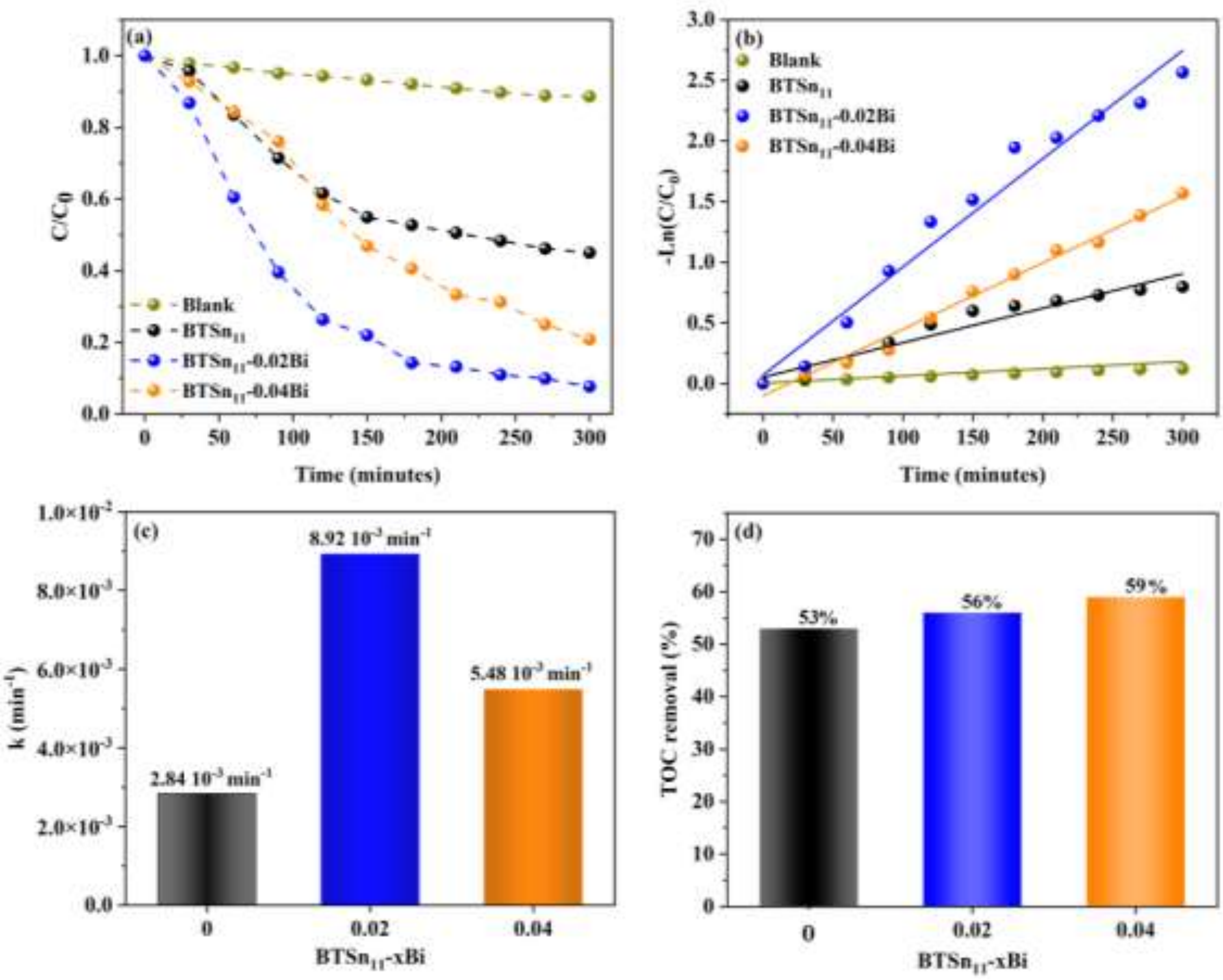

Fig. 5. (a) variation of the relative RhB concentration vs. vibration time using $BTSn_{11}$-xBi samples, (b) plots of $-\ln(C/C_0)$ vs. vibration time for the studied piezocatalysts, (c) the corresponding rate constant of each sample, (d) Proportion of Total Organic Carbon (TOC) removal analyzed upon RhB piezocatalytic degradation under low ultrasonic vibration energy.

Figures 5(a)–(c) show the kinetic rate of RhB degradation by $BTSn_{11}$-xBi powders under 30 W, 35 kHz ultrasonic vibration conditions. The rate constant is $8.92\times10^{-3}$ $min^{-1}$ for $BTSn_{11}$-0.02Bi, which is higher than those of $2.84\times10^{-3}$ $min^{-1}$ for pristine $BTSn_{11}$ and $5.48\times10^{-3}$ $min^{-1}$ for $BTSn_{11}$-0.04Bi sample. $BaZr_{0.02}Ti_{0.98}O_3$ powders demonstrated a rate degradation of RhB dye of $5.35\times10^{-3}$ $min^{-1}$ when subjected to 40 kHz and 70 W ultrasonic vibrations [51]. Our results confirm the highest piezocatalytic performance of the $BTSn_{11}$-0.02Bi sample for efficiently decomposing RhB dye under ultrasonic vibrations of 35 kHz and 30 W. The degradation performance of RhB using $BTSn_{11}$-0.02Bi is consistent with the piezoelectric current density response, confirming the remarkable piezocatalytic capability of this sample. The decrease in the efficiency of the piezocatalytic degradation of RhB becomes noticeable when the $Bi^{3+}$ doping level exceeds 2%. This decrease can be attributed to the formation of meaningful defects due to excessive doping, which can serve as complex sites for electron-hole [52]. Similarly, Chen et al. [53] demonstrated that when the carbon content was 2% in $BaTiO_3$/Carbon hybrid nanocomposites, the piezocatalytic degradation performance of RhB reached its maximum and decreased with increasing carbon content. Similarly, the piezocatalytic activity of RhB degradation using Ce-doped $BaTiO_3$ also showed a significant increase compared to the undoped sample [54]. In addition, the 3% Bi-doped $AgNbO_3$ catalyst demonstrated improved piezo-photocatalytic degradation of organic dyes compared to pristine $AgNbO_3$, revealing the significant role of Bi doping in boosting the catalytic performances [19]. Table 2 shows a clear comparison between the piezocatalytic performance towards RhB degradation by $BTSn_{11}$-0.02Bi and other perovskite-type systems.

Table 2. Comparison of the piezocatalytic performance of different perovskite materials for pollutant degradation.

| Piezocatalysts | Dye pollutant | Ultrasonic vibration | Kinetic rate constants (× $10^{-3}$) | Reference |
|---|---|---|---|---|
| $BTSn_{11}$-0.02Bi submicrocubes | RhB (5 mg/L) | 35 kHz, 30 W | 8.92 $min^{-1}$ | This work |
| $0.5Ba(Zr_{0.2}Ti_{0.8})O_3$-$0.5(Ba_{0.7}Sr_{0.3})TiO_3$ ceramic | MB (5 mg/L) | 40 kHz, 150 W | 4.3 $min^{-1}$ | [11] |
| $0.5Ba(Zr_{0.2}Ti_{0.8})O_3$-$0.5(Ba_{0.7}Ca_{0.3})TiO_3$ fibers | RhB (5 mg/L) | _ | 923.4 $h^{-1}$ | [55] |
| $(Ba,Sr)TiO_3$ nanowires | MO (5mg/L) | 40 kHz, 80 W | 19.6 $min^{-1}$ | [56] |
| $BaZr_{0.02}Ti_{0.98}O_3$ powders | RhB (6 mg/L) | 40 kHz, 70 W | 5.35 $min^{-1}$ | [51] |
| $(Ba_{0.85}Ca_{0.15})(Ti_{0.9}Zr_{0.1})O_3$ nanowires | MO (5 mg/L) | 40 kHz, 120 W | 7.1 $min^{-1}$ | [57] |

The mineralization of RhB was also investigated to evaluate its state of conversion to $H_2O$ and $CO_2$ upon piezocatalytic degradation. Fig. 5(d) shows the Total Organic Carbon (TOC) removal analyzed for the centrifuged dye solution at the end of the catalytic reaction using each piezocatalyst. The three piezocatalysts show a TOC removal higher than 50%. These results confirmed the excellent mineralization of RhB using $BTSn_{11}$-xBi under low-power ultrasonic vibration. In addition, total mineralization of RHB can be achieved if a longer time is devoted to the catalytic reaction.

Apart from the doping effect, most piezocatalytic systems utilize high-power ultrasonic vibration to trigger the dye decomposition process. Lan et al. [10] suggested that higher ultrasonic power or frequency can boost the piezo-photocatalytic performance of La-doped $BiFeO_3$ by increasing catalyst stress, thereby enhancing the piezoelectric field generated. Furthermore, Shi et al. [58] investigated the effect of the ultrasonic power (60, 90, 120, 150 W) at a constant frequency of 40 kHz on the degradation of RhB dye using $Na_{0.5}Bi_{0.5}TiO_3$ material. As expected, the RhB degradation rate reached its maximum of about 92% at the higher ultrasonic power of 150 W. Although high ultrasonic energy can improve piezocatalytic activity, it systematically increases energy consumption.

Moreover, high-vibration energy has several side effects, including excessive heating, high operating costs, and material deterioration. Therefore, exploring new piezocatalytic systems that harness low ultrasonic frequencies or power has become a hot topic due to the current energy crisis. For instance, Liu et al. [59] have successfully developed a piezocatalytic system to harvest low-frequency hydromechanical energy for piezocatalytic water disinfection applications. Consequently, using a low ultrasonic frequency can still induce the required piezoelectric effect while minimizing energy consumption. The reduction in vibration power is also of great interest. For this purpose developing new catalysts able to harvest low-power ultrasonic vibrations for waste water treatment is crucial. The remarkable performance of $BTSn_{11}$-0.02Bi makes it the best choice for RhB piezocatalytic degradation by harvesting low-power ultrasonic vibration.

### 3.5. Piezocatalytic reusability and mechanism for RhB degradation

The cyclic reusability test presented in Fig. 6(a) confirmed the remarkable stability of the $BTSn_{11}$-0.02Bi sample in the piezocatalytic degradation of the RhB dye under identical catalytic conditions. Even after three cycles, this sample showed virtually unchanged degradation efficiency.

In order to establish the piezocatalytic mechanism, charge trapping experiments were conducted to determine the main active species involved in the piezocatalytic degradation of RhB using $BTSn_{11}$-0.02Bi under ultrasonic vibration. This process follows the same process as the piezocatalytic evaluation, except that radical scavengers are involved. The test covers the detection of hydroxyl radicals ($^\bullet OH$), superoxide radical anions ($^\bullet O_2^-$), and piezo-induced holes ($h^+$). As illustrated in Fig. 6(b), the degradation efficiency of RhB was considerably reduced and dropped to a minimum when IPA was added during 300 minutes of ultrasonic vibration. Under the same experiment conditions, adding methanol as holes ($h^+$) scavenger also demonstrated an observable inhibition of the piezocatalytic degradation of RhB. In addition, compared to IPA and methanol, the introduction of $N_2$ as an $^\bullet O_2^-$ scavenger caused the smallest reduction in the piezocatalytic degradation of the RhB dye. Overall, the experiments involving the trapping of active species demonstrated that the piezocatalytic degradation of RhB appeared to be provided by piezo-induced hydroxyl radicals ($^\bullet OH$), holes ($h^+$), and superoxide ($^\bullet O_2^-$) radicals.

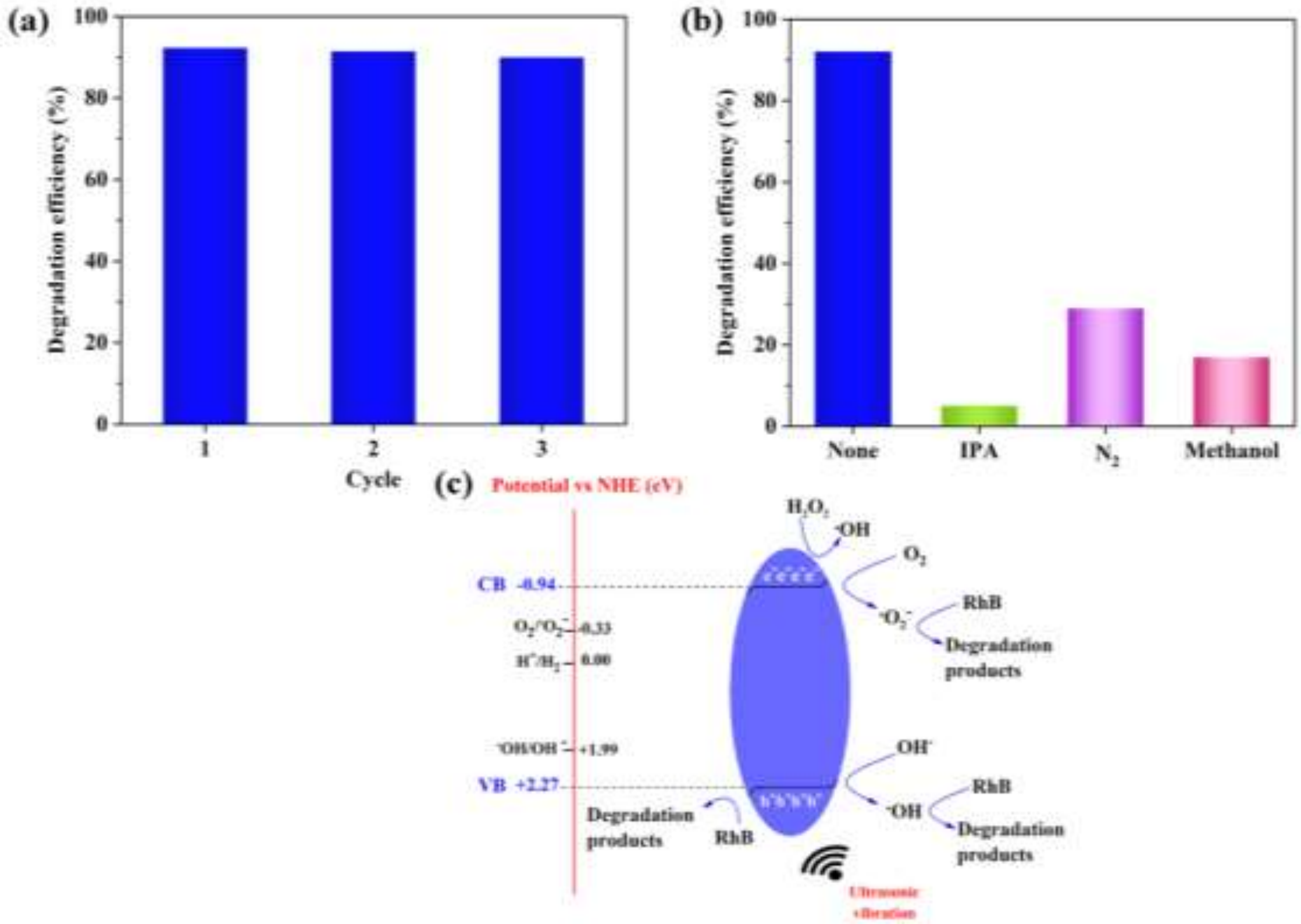

Fig. 6. (a) piezocatalytic degradation of RhB by $BTSn_{11}$-0.02Bi under ultrasonic vibration in the presence of scavengers, (b) cycling runs for RhB piezo-degradation using $BTSn_{11}$-0.02Bi catalyst, (c) schematic illustration of the possible piezocatalytic mechanism of RhB degradation using $BTSn_{11}$-0.02Bi catalyst.

According to the above measurement results, we have proposed the possible mechanism of the piezocatalytic degradation of RhB dye using the $BTSn_{11}$-0.02Bi catalyst (Fig. 6(c)). When the piezoelectric material is exposed to an external stress supplied by ultrasound, highly active bubbles can be produced, leading to the structural deformation of the piezoelectric catalyst [60]. This deformation leads to the generation of polarization charges, which in turn leads to the formation of a built-in electric field [61]. Otherwise, the lattice distortion caused by adding 2% of $Bi^{3+}$ ions in the A site of $BTSn_{11}$ makes $BTSn_{11}$-0.02Bi more sensitive to vibrational energy. The stronger piezoelectric field created by this distortion will enhance the efficient separation of electrons and holes within the piezoelectric catalyst. More accurately, the ultrasonic vibration initiates the piezocatalytic reaction, which generates electrons and holes (Equation 1). Throughout the piezocatalytic process, given that the CB potential of $BTSn_{11}$-0.02Bi (-0.942 eV) is more negative than $O_2/{}^{\bullet}O_2^{-}$ potential (-0.33 eV), $O_2$ can be reduced by the generated piezo-electrons ${}^{\bullet}O_2^{-}$ (Equation 2).

On the other hand, the $H_2O_2$ present in the medium can also react with the generated electrons to produce ${}^{\bullet}OH$ and $OH^-$ (Equation 3). Meanwhile, the holes ($h^+$) generated piezocatalytically can cause the RhB dye oxidation immediately and also react with $H_2O$ and/or $OH^-$ anions to generate ${}^{\bullet}OH$ radicals, which are capable of oxidizing RhB dye (Equations 4,5) [62]. Consequently, introducing an $^{\bullet}OH$ quencher significantly inhibits the catalytic process, in line with the findings of radical scavenging tests. The possible reactions involved in the piezocatalytic degradation of RhB dye are described below:

$$BTSn_{11}\text{-}0.02Bi + \text{vibration} \rightarrow h^+ + e^- \quad (1)$$

$$O_2 + e^- \rightarrow {}^{\bullet}O_2^- \quad (2)$$

$$H_2O_2 + e^- \rightarrow {}^{\bullet}OH + OH^- \quad (3)$$

$$OH^- + h^+ \rightarrow {}^{\bullet}OH \quad (4)$$

$$H_2O + h^+ \rightarrow {}^{\bullet}OH + H^+ \quad (5)$$

$$RhB + h^+ \rightarrow \text{Degraded products} \quad (6)$$

$$RhB + {}^{\bullet}OH \rightarrow \text{Degraded products} \quad (7)$$

$$RhB + {}^{\bullet}O_2^- \rightarrow \text{Degraded products} \quad (8)$$

## 4. Conclusions

The lead-free ferroelectric $BTSn_{11}$-xBi (x = 0, 0.02, 0.04) samples were successfully synthesized by the sol-gel method. Structural characterizations, morphological properties, bandgap evolution, piezoelectric current measurements, and piezoelectric properties were studied. The piezocatalytic activity of each sample was investigated through the degradation of RhB dye solution under the application of low-power ultrasonic vibration. The RhB degradation was only 54.94% for pristine $BTSn_{11}$ submicrocubes, while it increased for $BTSn_{11}$-0.02Bi, reaching its maximum of about 92%, and then decreased to 79% in the presence of $BTSn_{11}$-0.04Bi powder. The $BTSn_{11}$-0.02Bi powder exhibited the lowest band gap, smallest particle size, and strongest piezoelectric current. Moreover, its PFM hysteresis loops required the lowest applied field. TOC measurements revealed good mineralization of the RhB dye using $BTSn_{11}$-xBi piezocatalysts. The degradation efficiency of the $BTSn_{11}$-0.02Bi sample manifested only a slight reduction after 3 recycling tests, revealing its high stability. Active species trapping experiments demonstrated that $^{\bullet}OH$, $h^+$, and $^{\bullet}O_2^-$ significantly contributed to the RhB degradation process using $BTSn_{11}$-0.02Bi. The possible piezocatalytic mechanism has been suggested based on the results of the above experiments. The results of this study show that $Bi^{3+}$ ion doping can effectively enhance the piezocatalytic performance of $BTSn_{11}$ by applying only low ultrasonic vibration energy supported by a hydroxyl radical source. This synergistic approach addressed the limitations of ultrasonic vibration and considerably enhanced its performance, significantly improving overall process efficiency.

**Acknowledgments**

This work was supported by HORIZON-MSCA-2022-SE H-GREEN (No. 101130520), MSCA-2020-RISE-MELON (No. 872631) and the Region of Hauts-De-France (HDF). The Slovenian Research and Innovation Agency (research core funding P2-0105, P1-0125 and research project N2-0212) is acknowledged. H. U. thanks Val Fišinger, Jena Cilenšek, and Katarina Lubina (Erasmus+ programme) for help in the laboratory.